\documentclass[twocolumn,showpacs,preprintnumbers,amsmath,amssymb]{revtex4}

\usepackage{graphicx}
\usepackage{dcolumn}
\usepackage{bm}

\def\Journal#1#2#3#4{{#1} {\bf#2}, #3 (#4)}

\def\PRL{Phys. Rev. Lett.}
\def\PRD{Phys. Rev. D}
\def\NIMA{Nucl. Instr. Meth. A}
\def\JPG{J. Phys. G}

\def\meanY{$\left<y\right>$}
\newcommand{\be}{\begin{equation}}
\newcommand{\ee}{\end{equation}}
\newcommand{\bea}{\begin{eqnarray}}
\newcommand{\eea}{\end{eqnarray}}

\begin{document}

\title{$\Lambda$ Production at High Rapidity in d+Au Collisions at 
\boldmath$\sqrt{s_\mathrm{NN}} = 200$\unboldmath\,GeV}
\author{Frank Simon}
\email{fsimon@mppmu.mpg.de}
\affiliation{%
(for the STAR Collaboration)\\
\\
Max--Planck--Institut f\"ur Physik\\
F\"ohringer Ring 6\\
80805 M\"unchen\\
Germany}%

\date{\today}

\begin{abstract}
We present first preliminary studies of $\Lambda$ and $\overline{\Lambda}$ 
production
in the pseudorapidity region $2.5 < |\eta| < 4$, covered by the forward
radial-drift TPCs (FTPCs) in STAR. The FTPCs provide momentum and charge
determination but no particle identification, making the use of combinatorial
methods and background subtraction necessary for $\Lambda$ identification.

The $\overline{\Lambda}/\Lambda$ ratio measured at high rapidity is 
compared to the ratio
obtained at mid-rapidity with the STAR TPC. Differences in the ratio at 
positive and
negative rapidity point to an asymmetry in particle and antiparticle 
production in
d+Au collisions.

These results have been presented as a poster at Quark Matter 2004 in 
Oakland, California.
\end{abstract}

\maketitle

\section{Introduction}

In relativistic heavy ion collisions at the highest available RHIC energies,
the ratio of yields of antiparticles to particles reaches the highest
values yet observed. Even the baryon ratios $\overline{p}/p$ and
$\overline{\Lambda}/\Lambda$ reach values near unity, indicating an
almost net baryon free environment at mid--rapidity. The antiparticle to
particle ratios are observed to be flat as a function of rapidity ($y$) close
to mid--rapidity \cite{BrahmsAuAu, StarAuAu}.

Away from mid--rapidity, the baryon content of the beam nuclei comes into
play, and, in addition to particle--antiparticle pair production, other
processes contribute significantly to the particle production.
Measurements by the BRAHMS collaboration \cite{BrahmsAuAu} show a
significant drop of $\overline{p}/p$ in Au+Au collisions starting at 
$y \sim 1$.
In asymmetric collision systems, measurements at lower energies by 
NA49 with p+Pb collisions
show different contributions of baryon number transfer for projectile 
and target 
rapidity regions due to multiple collisions suffered by the projectile 
nucleon, but not by the target \cite{Rybicki}.

In the present paper, first preliminary measurements of 
$\overline{\Lambda}/\Lambda$ at $|y| \sim 2.7$ in d+Au 
collisions are presented.

\section{Detector and Analysis Technique}

\begin{figure}
	\includegraphics[width=80mm]{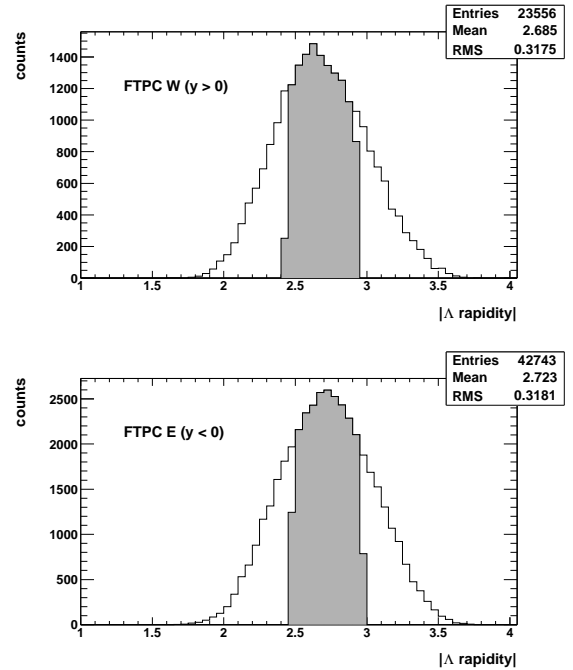}
	\caption{Rapidity of $\Lambda$ candidates 
(including $K^0_s$ background) for both FTPCs. 
	The rapidity range used for the present analysis is shown as 
the shaded histogram.}
	\label{fig:LambdaRapidity}
\end{figure}

The two radial--drift forward time projection chambers (FTPCs) \cite{Ftpc}
of the STAR \cite{Star} experiment permit the study of charged hadrons 
at forward
rapidity in heavy ion collisions. This extends the acceptance of the
spectrometer towards the fragmentation region and gives access to phenomena
away from mid--rapidity. However, the FTPCs measure a maximum of 10 hits per
track, which makes particle identification via energy loss measurements
impossible with the current state of detector calibrations.

\begin{figure*}
	\includegraphics[width=179mm]{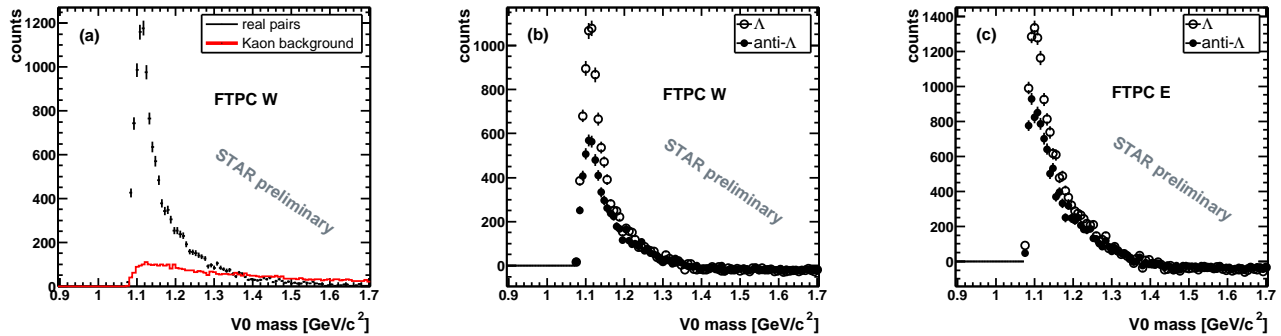}

	\caption{a) Invariant mass distribution of $\Lambda$ candidates 
in FTPC W (deuteron side) 
	with estimated $K^0_s$ background distribution, b) and c) show the 
background subtracted
	$\Lambda$ and $\overline{\Lambda}$ invariant mass distributions. 
The widths of the peaks are due 
	to the limited momentum resolution of the detectors, and can be 
reproduced by simulations.}
	\label{fig:LambdaInvariant}
\end{figure*}

The two FTPCs west (positive rapidity) and east (negative rapidity)
of the interaction point in STAR are well
located to study asymmetries in d+Au collisions. In the 2003 d+Au beam time,
the deuterons were entering STAR from the east, gold from the west.
Hence the particle multiplicity in FTPC east is higher than in west,
since particles produced from the gold nucleons preferentially continue 
towards the east.

Since the measurable decay mode of the 
$\Lambda$ is $\Lambda \rightarrow p\, \pi$
(with a branching ratio of 64\%), the lack of particle identification 
makes the use
of all positive particles as $p$ candidates for the $\Lambda$ case or 
all negative particles as $\overline{p}$ for the $\overline{\Lambda}$ case
necessary. This introduces a considerable background to the measurement.
Strict geometric cuts on the assumed
daughter tracks  and the resulting $\Lambda$ candidate help to reduce the
background. The most important cuts were on the distance of closest
approach (dca) to the primary vertex of the daughters, which should be 
relatively 
small for the $p$ candidate and large for the $\pi$ candidate,
and cuts on the dca and the decay length of the resulting $\Lambda$ candidate.

 The major source of background remaining after these cuts is
estimated to be from $K^{0}_{s}\, \rightarrow \, \pi^+\pi^-$, where one of 
the two daughter
pions is assumed to be a proton. For the current analysis a full 
GEANT detector simulation with a HIJING \cite{HIJING} generated $K^0_s$ 
distribution in 
$p_t$ and $y$ was used to predicted this background produced by making the 
wrong mass assumption for one of the two daughters.

Although the analysis presented
here does not correct the yields for acceptance and efficiency, simulation
studies show that the corrections for $\Lambda$ and
$\overline{\Lambda}$ are equal to first order. This permits the calculation 
of antiparticle
to particle ratios without the knowledge of the absolute yield. The resulting
ratios are not corrected for absorption or annihilation of the $\Lambda$ or 
its daughter
particles in the detector material. Due to the high momentum of the particles 
in forward
direction, absorption effects are small. Simulations show them to be less 
than 2\%,
which is of the same order as the statistical error of the study.

\section{Analysis Results}

For the current analysis, a sample of 10.6 million d+Au minimum bias events 
with a reconstructed primary vertex within 50 cm of the nominal interaction 
point are used.

With the assumption that all candidates that pass the cuts are actually 
$\Lambda$s, their rapidity can be calculated. This still includes all 
background that remains 
after the geometrical cuts. Figure \ref{fig:LambdaRapidity} shows the total
rapidity acceptance of the analysis for both detectors and the used 
range within $\pm$ 0.25 of the mean rapidity \meanY. 

The rapidity slice used for the 
analysis is limited to 0.5 units because
HIJING simulations predict a strong rapidity dependence of 
$\overline{\Lambda}/\Lambda$ in the region of interest, which would affect the 
measurement if a wide range of rapidities is included in one bin. In addition, 
the $p_t$ acceptance of the FTPCs is constant as a function of the 
pseudorapidity $\eta$, not as a function of $y$. In a wide analysis window in 
$y$ the covered $p_t$ changes considerably over the
selected $y$ range, and a possible $p_t$ dependence of the ratio would make 
the interpretation of the results more difficult, especially if this 
dependence is different for the deuteron and the gold side of the collision. 

Figure \ref{fig:LambdaInvariant} a) shows in black the invariant mass 
distribution of all positive--negative pairs with 
0.4 GeV/c $< p_t <$ 1.5 GeV/c that 
fulfill the geometric cuts in FTPC west with the assumption that 
positive tracks are protons and
negative tracks are pions.
The red curve is the estimated background from wrongly identified Kaons, 
which is normalized
to the integral of the invariant mass distribution in the range 
from 1.3 GeV/c$^2$ to 2.0 GeV/c$^2$.

This background is subtracted bin by bin and leads to the $\Lambda$ and 
$\overline{\Lambda}$ invariant
mass distributions shown for both detectors in figure 
\ref{fig:LambdaInvariant} b) and \ref{fig:LambdaInvariant} c). 
The widths of the peaks are in agreement with simulations and due to the 
limited momentum resolution of 
the FTPCs, especially for tracks not originating from the primary vertex. 
From these invariant mass distributions the uncorrected yields and the 
statistical errors for 
$\Lambda$ and $\overline{\Lambda}$ at forward and backward rapidity are 
determined by summing all bins 1.08 GeV/c$^2 < M_{inv} <$ 1.18 GeV/c$^2$.

The systematic errors have not been 
investigated in detail yet. From a variation of cuts and background 
estimations 
they are predicted to be 0.12 for this preliminary analysis. Since
possible remaining background contributions tend to be equal for $\Lambda$ and
$\overline{\Lambda}$, an asymmetry of the systematic errors towards lower 
values
of the ratio appears likely, but has not yet been thoroughly investigated.
The derived antiparticle to particle ratios
are compared with the ratios given by the HIJING event generator.

The mean rapidity of $\Lambda$ and $\overline{\Lambda}$
candidates on the deuteron side (FTPC W) is \meanY\ = 2.69. The analysis 
yields a ratio of
$\overline{\Lambda}/\Lambda = 0.58 \pm 0.02$(stat)$\pm 0.12$(syst).
On the gold side (FTPC E) \meanY\ = -2.72, with a ratio
$\overline{\Lambda}/\Lambda = 0.71 \pm 0.02$(stat)$\pm 0.12$(syst).

\begin{figure}
	\includegraphics[width=80mm]{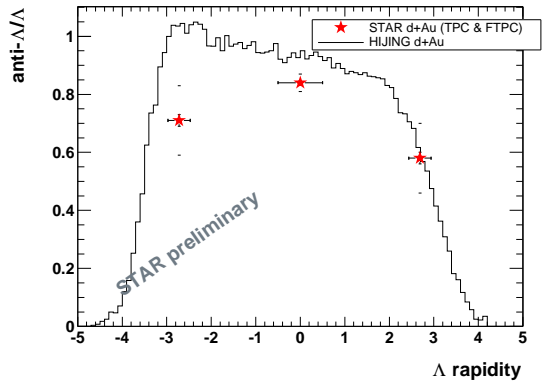}
	\caption{$\overline{\Lambda}/\Lambda$ ratio as a function of rapidity.
	The three STAR points are compared with HIJING calculations. 
The statistical 
	errors are shown as the vertical bars, systematic errors 
(estimated to be 0.12 in the FTPCs) 
	are indicated by the small horizontal lines. The horizontal error 
bars show the bin width 
	in $y$ used to extract the ratio}
	\label{fig:Ratio}
\end{figure}

Figure \ref{fig:Ratio} shows $\overline{\Lambda}/\Lambda$ from HIJING
without taking detector effects into account. It has been shown by measurements
in Au+Au collisions at $\sqrt{s_{NN}} = 130$ GeV \cite{Phenix, Brahms130} 
that the pure HIJING model tends to underestimate 
the influence of baryon number transport, and thus overestimates the 
antibaryon/baryon ratio.
Overlaid in red are the STAR points, the mid--rapidity value being 0.84,
determined from a preliminary analysis of main TPC data. The data show good 
agreement 
with HIJING on the deuteron side, while on the gold side, HIJING appears to 
overestimate the ratio.

$\overline{\Lambda}/\Lambda$ shows a significant drop at high rapidity, 
probably due to baryon number conservation
and fragmentation contributions to the $\Lambda$ production. The 
asymmetry of the ratio in 
the highly asymmetric collision system may be caused by different 
contributions of pair production 
and baryon number transport in the projectile and the target region.
While the participating nucleons in the gold nucleus typically 
suffer only a single collision each, the nucleons from the deuteron 
participate in multiple
collisions as they pass through the gold nucleus. This can lead to increased 
baryon number transport
from the deuteron region towards mid--rapidity and thus results in a 
decrease of 
$\overline{\Lambda}/\Lambda$ in the studied rapidity range. 

\section{Conclusion}

First preliminary measurements of $\overline{\Lambda}/\Lambda$ at 
high rapidity in
d+Au collisions at $\sqrt{s_{NN}}$ = 200 GeV using the forward TPCs of the STAR
experiment have been presented. An asymmetry between
the deuteron and the gold side of the collision is seen, indicating different
contributions of antiparticle and particle production mechanisms and 
baryon number
transport in the forward region. In general, the ratios at high rapidity 
are lower than at mid--rapidity.

Future studies will compare to a wider range of theoretical models and 
use a variety of background models and cut sets.  
A binning of the analysis in collision centrality, rapidity or $p_{t}$ 
might also be feasible.

\end{document}